\def\be{\begin{equation}}
\def\ee{\end{equation}}

\documentclass[12pt]{article}
\usepackage{amsmath}
\usepackage{amssymb}
\title{Local Lagrangian for Exponentially Large Extra Dimensions} 
\author{\large 
\textbf{Michael J. May\footnote{email: mjmay@pha.jhu.edu}
\mbox{  }and Raman Sundrum\footnote{email: sundrum@pha.jhu.edu}}\\
\emph{
Department of Physics and Astronomy} \\ 
\emph{Johns Hopkins University} \\ 
\emph{3400 North Charles St}. \\ 
\emph{Baltimore, MD 21218-2686}}

\begin{document}
\maketitle

\begin{abstract}

We write an explicit local action for a large extra dimensions
 stabilization scenario due to Arkani-Hamed, Hall, Smith and Weiner (AHSW).  Our action allows the AHSW proposal to be generalized to non-Poincar\'{e} invariant configurations, supersymmetric extensions, quantum effective field theory, and cosmological scenarios. The central step in constructing the action is working in terms of the four form gauge field which is the ``electric-magnetic'' dual to the ``magnetic'' scalar of AHSW. The action is manifestly invariant under
higher general coordinate invariance.
\end{abstract}

\section{Introduction}

The issue of why the 4D Planck scale is so much larger than the weak scale remains an open question. Models based on extra dimensions have attempted to solve this hierarchy problem. 
Arkani-Hamed,  Dimopoulos and Dvali (ADD) \cite{ADD1, ADD2} 
proposed a scheme whereby two, compact, large extra dimensions can be used to lower the six-dimensional Planck scale, $M^{(6)}_{Pl}$, to near the weak scale while having a large four-dimensional Planck scale, $M^{(4)}_{Pl}$.    

An essential ingredient for the internal consistency, phenomenology, and cosmology of these types of models is natural stabilization of the large radion VEV.
Arkani-Hamed, Hall, Smith and Weiner
(AHSW) have propopsed  a classical 
mechanism to stabilize co-dimension two ADD models which yields an effective potential for the radion field, $a$, of the form \cite{AHSW}:
\be
\label{addexp}
 V(a)= \frac{c_1}{\ln(a / \delta)}+c_2 \ln(a / \delta) + V_c,
\ee
where $c_1$ and $c_2$ are constants, $V_c$ is a four-dimensional cosmological constant term, and $\delta \gtrsim (M^{(6)}_{Pl})^{-1}$ is a cutoff.  This logrithmic dependence on $a$ can naturally lead
 to a radion VEV which is  exponentially larger than $\delta$,
\be
\label{expvev}
<a>= \delta e^{\sqrt{\frac{c_1}{c_2}}}.
\ee

AHSW stabilization is certainly attractive. However, the derivation 
relies on a ``magnetic'' boundary condition for one of the scalar fields, which 
cannot be reproduced by any local Lagrangian interactions for the scalar. This 
makes it difficult to understand general processes outside the 
4D Poincar\'{e} ansatz (which simplifies the vacuum stabilization analysis), as well 
as quantum processes. In this paper, we present a local action with manifest 
six-dimensional general coordinate invariance which reproduces the AHSW 
stabilization scenario. The key is to work in terms of a four-form gauge field which 
is the ``electric-magnetic'' dual to the AHSW scalar.  The action we are proposing is inspired by a similar situation arising in cosmic strings \cite{ewcosstr, vv, DS88}. For a review see Ref. \cite{cosmicstrings}.
  
  Our action has several applications.  It may serve as the basis for a supersymmetric version of ADD which is
 needed to keep the extra dimensions naturally flat.  It can be treated as a quantum effective field theory, and it can be applied to cosmological scenarios where extra-dimensional moduli dynamics play an important role \cite{gravpotstr, shapequint}.  It is also possible that aspects of AHSW stabilization, and our local 
description of it, will be useful in different contexts from ADD.



In section \ref{localaction}, we present our action and explain how we were led to it.  In section \ref{newstuff}, we derive the equations of motion that lead to Eq. (\ref{addexp}) from our action using the same simplifying assumptions as in Ref. \cite{AHSW}.

\section{A Local Action}
\label{localaction}

One would like to write down an effective field theory action where the interactions between the branes and the bulk fields are local and explicit.  The main difficulty is that in the AHSW scenario one bulk field, call it $\phi_2$, is the phase of a complex scalar which is taken to wind around two of the three branes in the ADD model.   This is analagous to the phase of a complex scalar about a global $U(1)$ string defect \cite{ewcosstr, vv, DS88}. The scale, $\delta$,  can be interpreted as the scale at which the vortex interior can be resolved.  The interior structure is in the radial profile of the complex scalar.  It is impossible, without specifying the interior structure of the brane, to write down a local interaction reproducing the winding boundary conditions in terms of only light degrees of freedom.  Here, the light degrees of freedom are the ``magnetic'' scalar, $\phi_2$, and the brane fluctuations, $Y^M_{(n)}$. (We use the notation developed in Refs. \cite{eft3br, cmp3br} to discuss the brane fluctuations.)  The inability to write down such a local brane interaction is not a handicap when strictly considering a four-dimensional Poincar\'{e} invariant configuration.   It does pose a problem when general processes are considered.  Here we suggest an effective field theory where the branes directly couple to the ``electric-magnetic'' dual to $\phi_2$ so as to reproduce the winding boundary condition in a local way.  

We start with the action $S$. The six-dimensional fields appearing in our action are a conventional scalar $\phi_1$, a four-form gauge field $B_{LMNP}$, the six-dimensional metric $G_{MN}$, and the embedding of three 3-branes $Y^M_{(n)}(x)$ which are labeled by $n=1,2,3$.  Given the anti-symmetry of $B_{LMNP}$, we can write the anti-symmetric field strength by permuting indicies, 
\be
H^{NLPQR} = \partial^N B^{LPQR} + \partial^R B^{NLPQ} + \partial^Q B^{RNLP}+
\partial^P B^{QRNL}+ \partial^L B^{PQRN}.
\ee
Then, the scalar of AHSW is the ``electric-magnetic'' dual of $B_{MNLP}$. 
\be
\label{dual1}
\partial_M \phi_2 = \frac{w^2}{5!} \epsilon_{MNLPQR} \: H^{NLPQR},
\ee
where $w$ is a constant with mass dimension one. We take all mass scales to be set by the six-dimensional Planck scale, $M^{(6)}_{Pl} \sim$ TeV. 

The action is
\be
\label{action}
\begin{split}
S=&S_{grav}+S_{\phi}+S_{B} \\
S_{grav}=&- \int d^6X \sqrt{-G} \,2 \, (M^{(6)}_{Pl})^4 R^{(6)} - 
\sum_{n=1}^{3} \int d^4x \sqrt{-g_{(n)}} f^{4}_{(n)} \\
S_\phi=&   \int d^6X \sqrt{-G} \; \frac{ G^{MN}\partial_M \phi_1 \partial_N \phi_1}{2} + \sum_{n=1}^{2} \int d^4x \sqrt{-g_{(n)}} \;  \lambda (\phi_1(Y_{(n)}) +(-1)^n \nu^2)^2 
 \\
S_{B} =& \int d^6X \sqrt{-G} \; \frac{-w^4}{10} H_{MNLPQ}H^{MNLPQ} + \sum_{n=1}^{2} (-1)^n q w^4  \int B_{MNLP} d \sigma_{(n)}^{MNLP}, \end{split}
\ee
where $\lambda$ is a dimensionless coupling, and $q$ is the quantum of brane 
charge.  For two branes, $q= \pm 1$.  For the third, $q=0$, indeed note that 
this brane
has no direct interactions with stabilizing fields, only with gravity.   
$\nu$ is a constant with mass dimension one that parameterizes brane tadpoles for $\phi_1$. The $f^4_{(n)}$ are the brane tensions.  We define 
\be
 g^{(n)}_{\mu \nu}(x) \equiv G_{MN} \big( Y_{(n)}(x) \big) \, \partial_\mu Y_{(n)}^M \, \partial_\nu Y_{(n)}^N,
\ee
the induced metrics on each brane, and also
\be
 d \sigma^{MNLP}_{(n)} \equiv \epsilon^{\mu \nu \alpha \beta} \partial_\mu Y^M_{(n)} \partial_\nu Y^N_{(n)}
\partial_\alpha Y^L_{(n)} \partial_\beta Y^P_{(n)} d^4x,
\ee
the induced measure on each brane.  Note that the tensor $\epsilon^{\mu \nu \alpha \beta} / \sqrt{-g_{(n)}}$ has cancelled the metric determinant in the 
last term of $S_B$.

The $1/ \ln(a/\delta)$ term in the effective potential, Eq. (\ref{addexp}), follows easily by integrating out $\phi_1$ from this action as in AHSW for sufficiently large $\nu^2$.  The term proportional to $\ln(a/\delta)$ arises from integrating out the AHSW ``magnetic'' scalar. We will show that integrating out $B^{MNLP}$ from our action using the same simplifying 
approximations made by AHSW also leads to Eq. (\ref{addexp}).

\section{Effective Potential from Dual Description}

\label{newstuff}

We solve for the bulk stabilizing fields in the background geometry induced 
predominantly by the branes.   The gravitational back-reaction of the stabilizing bulk fields can self-consistently be taken to be small.  The action $S_{grav}$ describes three 3-branes with tensions $f^4_{(n)}$ in 6-dimensional space-time.  Adopting the four-dimensional Poincar\'{e} invariant ansatz,
\be
\label{metric}
ds^2=\eta_{\mu \nu} dx^\mu dx^\nu + G_{mn} (X^4, X^5) dX^m dX^n,
\ee
the extra dimensions mimic static gravity in two spatial-dimensions; the branes appear as point sources in the two dimensional plane spanned by the extra dimensions \cite{grav2+1, 3dgrav}. We will focus on the extra two dimensions where space-time is flat everywhere except at the conical singularities which coincide with the brane locations. The conical deficit angles, $\alpha_n$, are determined by the brane tensions,
\be
\alpha_n = \frac{f^4_{(n)}}{4 M^4}.
\ee
To obtain the static, compact solution, the Gauss-Bonnet theorem requires 
\be
\sum_{n=1}^3 \alpha_n = 4 \pi.
\ee

The $B_{LMPQ}$ equation of motion is
\be
\partial_K(\sqrt{-G} H^{KLMNP})=  - q \int d \sigma^{LMNP}_{(1)} \; 
\delta^{(6)} \Big ( X-Y_1(x) \Big )  +  q \int d \sigma^{LMNP}_{(2)} \; 
\delta^{(6)} \Big ( X-Y_2(x) \Big ).
\ee
Applying the four dimensional Poincar\'{e} invariant ansatz, we can take (for $i=1,2$) 
\be
\label{g2}
Y_{(i)}^\mu=X^\mu \qquad Y_{(i)}^4=0 \qquad Y_{(i)}^5= \pm \frac{a}{2},
\ee
so that the only non-trivial equation of motion is
\be
\label{bc}
\partial_M \partial^M B^{0123}=  q \delta (X^4) \delta(X^5 - \frac{a}{2})- q \delta (X^4) \delta(X^5 +\frac{a}{2}).
\ee
$B^{0123}$ appears as a four dimensional scalar. In polar coordinates centered around the $n$\raisebox{0.5ex}{\footnotesize{th}} charged brane, Stokes' theorem provides a boundary condition,
\be
\label{qbc}
\lim_{ \rho \rightarrow 0} \int_0^{\theta} \frac{\partial B^{0123}}{\partial \rho} \; \rho  d \beta_{n} = (-1)^n q,
\ee 
where $\theta = 2 \pi - \alpha$.  
For simplicity we take $\alpha_{1,2}=\alpha$, that is, the first two branes have equal tension. $\beta_n$ is the natural angular coordinate from $0$ to 
$\theta$ around each of the charged branes.   

The solution to Eq. (\ref{bc}) in the compact space can be approximated from a non-compact one using the same assumptions as in Ref. \cite{AHSW}.  Near a particular brane the solution is dominated by that brane and is insensitive to the other branes. As a first approximation, Eq. (\ref{bc}) is solved in ``pie-wedge'' slices near each brane separately and the solutions are patched together. This yields, 
\be
\label{soln}
B^{0123} = \frac{q}{\theta} \big( \ln(\rho_+)-\ln(\rho_-) \big) + F,
\ee
where $\rho_+$ is the distance to the positively charged brane, and $\rho_-$ the distance to the negative brane.  The first and second terms give the behavior of $B^{0123}$ near the positively and negatively charged branes 
respectively (that is in the absence of the other branes), 
while $F$ gives the remaining behavior away from these branes.  $F$ patches the two independent solutions together and enforces the compactness of the boundary conditions.  Note, $F$ is dimensionless and smooth since the singularities 
from the charged branes, from satisfying Eq. (\ref{qbc}), 
are in the log terms of Eq. (\ref{soln}). $F$ is therefore insenstive to any short-distance cutoff for the structure of the branes.  

    In intgerating out $B_{LMNP}$, the field strength integral, 
$\int H^2 d^2X_m$, contains an integrand of the form
\be
(\partial_{\rho} \ln \rho)^2 \rho,
\ee
which contributes a log divergent piece to the AHSW effective potential, Eq. (\ref{addexp}).  The log is cutoff by the short distance physics of the brane, parameterized by $\delta$.  However, the smoothness of $F$ ensures any terms in $\int H^2 d^2X_m$ involving $F$ are insensitive to $\delta$. 
Therefore, the $F$ dependent piece of the field strength integral at most contributes to $V_c$ in Eq. (\ref{addexp}) plus terms that vanish in the limit 
of small $\delta$. 
Thus, adding the contribution from integrating out $\phi_1$ 
we have reproduced
 the AHSW effective potential of Eq. (\ref{addexp}).

\section{Acknowledgements}
The work of M. M. and R. S. 
was supported in part by NSF Grant P420D3620414350.  R. S. was also 
supported in part by  DOE Outstanding Junior Investigator Award DEFG0201ER41198.

\bibliography{my}
\bibliographystyle{h-physrev}

\end{document}